\documentclass[12pt]{article}
\usepackage{amssymb}
\usepackage{amsmath}
\usepackage{amsfonts}
\usepackage{amsthm}

\font\frak=eufm10 scaled\magstep1

\font\bigblack=msbm10 scaled\magstep 2
\font\bbigblack=msbm10 scaled\magstep3

\def\d{{d}}

\def\bigfield #1{\hbox{{\bigblack #1}}}
\def\bbigfield #1{\hbox{{\bbigblack #1}}}

\def\v #1{\vert #1\vert}             

\def\m #1 #2{(-1)^{{\v #1} {\v #2}}} 
\def\pd#1#2{\frac{\partial#1}{\partial#2}}

\def\<#1>{\langle#1\rangle}        
\def\>#1{{\bf #1}}                
\def\f(#1,#2){\frac{#1}{#2}}

\def\dt2#1{\frac{d^2 #1}{dt^2}}

\def\ea{\varepsilon_a}

\def\d{\delta}

\def\big R{{\hbox{{\bigfield R}}}}
\def\bbig R{{\hbox{{\bbigfield R}}}}
\def\N{{\hbox{{\field N}}}}         

\def\dim{\hbox{{\rm dim}}}        

\def\ba{\begin{eqnarray}}
\def\ea{\end{eqnarray}}
\def\be{\begin{equation}}
\def\ee{\end{equation}}

\def\d{\rm d}                  

\def\<#1>{\langle#1\rangle}

\def\pd#1#2{\frac{\partial#1}{\partial#2}}

\def\N{{\frak N}}

\def\di{\bigstar}
\newcommand{\bea}{\begin{eqnarray}}
\newcommand{\eea}{\end{eqnarray}}

\def\pd#1#2{\frac{\partial#1}{\partial#2}}

\theoremstyle{plain}
\newtheorem{theorem}{Theorem}

\newtheorem{proposition}{Proposition}
\newtheorem{definition}{Definition}
\newtheorem{lemma}{Lemma}

\def\N{{\frak N}}

\font\frak=eufm10 scaled\magstep1

\def\<#1>{\langle#1\rangle}

\begin{document}

\centerline{\Large {\bf Quasi-Lie schemes and Emden--Fowler equations}} \vskip 0.75cm

\centerline{ Jos\'e F. Cari\~nena$^{\dagger}$, P.G.L. Leach$^{\ddagger}$ and Javier de Lucas$^{\dagger,\S}$}
\vskip 0.5cm

\centerline{$^{\dagger}$Departamento de  F\'{\i}sica Te\'orica and IUMA, Universidad de Zaragoza,}
\medskip
\centerline{50009 Zaragoza, Spain.}
\medskip
\centerline{$^{\ddagger}$School of Mathematical Sciences, University of KwaZulu-Natal,}
\medskip
\centerline{Private Bag X54001 Durban 4000, Republic of South Africa.}
\medskip
\centerline{$^{\S}$Institute of Mathematics, Polish Academy of Sciences,}
\medskip
\centerline{ul. \'Sniadeckich 8, P.O. Box 21, 00-956, Warszawa, Poland.}
\medskip
\vskip 1cm

\begin{abstract}
The  recently developed  theory of quasi-Lie schemes
 is studied and applied to investigate several equations of Emden type and  a scheme to
 deal with them and some of their generalisations is given. As a first
result we obtain  $t$-dependent constants of the motion for
particular instances of Emden equations by means of some of their
particular solutions. Previously known results are recovered from
this new perspective. Some  $t$-dependent constants
 of the motion for equations of Emden type satisfying certain conditions are
 recovered. Finally new exact particular solutions are given for
 certain cases of Emden equations.
\end{abstract}

\section{Introduction.}
\indent 

Systems of nonautonomous first-order differential equations appear broadly in
Mathematics, Physics, Chemistry and Engineering. Therefore methods to
 solve  these systems and analyse their
properties are specially interesting because they allow us to
understand many important problems in these various fields.

As a first insight into this topic the theory of Lie systems can be considered
\cite{LS,PW,CGM07}. Many applications have recently been studied through this theory
\cite{CarRam,CLR07b}. Nevertheless there are many differential
equations which cannot be studied by means of the theory of Lie systems
 \cite{CGL08,CL08d}. To treat more general differential equations a new theory has been recently developed. This theory is based on the so-called
quasi-Lie schemes \cite{CGL08}. These schemes generalise the concept of Lie
system and sometimes they allow us to transform a system of differential
equations, generally a non-Lie system, into a Lie system. Once the final Lie
system is studied and its properties are found, the theory of quasi-Lie schemes
provides constants of the motion,  $t$-dependent superposition rules or even
solutions for the initial system of differential equations.

Quasi-Lie schemes have been useful to deal with many systems of differential
equations.
 They have been applied to study nonlinear oscillators \cite{CL08}, dissipative
 Ermakov systems \cite{CL08d,CL08}, etc. In this
 paper we apply quasi-Lie schemes to investigate the properties of
 Emden--Fowler equations. The literature about these equations is very large
 \cite{PK06}--\cite{Le85} and applications of these equations can be found, for
 example, in Mathematical Physics, Theoretical Physics, Astronomy,
 Astrophysics and Chemical Physics. We only consider
 a small sample and, for instance, there are about 140 references in the review by Wong in 1977 \cite {Wong76a}
 .

Our aim in this paper is to investigate the properties of
Emden--Fowler equations from the point of view of the theory of
quasi-Lie schemes and quasi-Lie systems. We firstly show that the
knowledge of certain particular solutions allows us to transform a
given Emden--Fowler equation into a Lie system by means of a
quasi-Lie scheme and determine a constant of the motion through
this particular solution. We also study the generalised
Emden--Fowler equation to recover from our point of view the
origin of the Kummer-Liouville transformation. Next, use is made
of the transformation properties of our quasi-Lie scheme in order
to obtain some $t$-dependent constants of the motion for families
of Emden type equations satisfying some conditions. Finally, we
derive a family of exact particular solutions for a particular
Emden-Fowler equation by means of a certain kind of $t$-dependent
superposition rule.

In this paper we start in Section 2 with a report on the theory of Lie systems and Lie schemes
 in order to review
 in Section 3 some previous results about quasi-Lie schemes and Emden
 equations and provide some new details. Section 4 is devoted to show that certain particular solutions of Emden equations enable us to obtain $t$-dependent constants of the
 motion.
 The latter result is applied in Section 5 to study some particular cases of Emden equations and
build up some particular instances of such equations for which we
can obtain a  $t$-dependent constant of the motion by means of our
method.  A generalised Emden-Fowler equation is studied in Section
6 and we get the Kummer-Liouville transformation from our
framework. In Section 7 we use the transformation properties of
our scheme to obtain $t$-dependent constants of the motion for
certain equations of Emden type satisfying some integrability
conditions. Finally, in Section 8 we apply some results obtained
along the paper to analyse certain Emden-Fowler equations. As a
result, we recover some results about these equations and we find
a $t$-dependent partial superposition rule. Such a superposition
rule is used next to obtain a family of solutions for an important particular
 Emden-Fowler equation.

\section{Review on Lie and quasi-Lie systems.}
\qquad In this Section we report some previous results on the theory of Lie systems, $t$-dependent vector fields and quasi-Lie systems and schemes. Furthermore we define the extended group of transformations in order to improve the methods of the theory of quasi-Lie schemes. Most results and mathematical objects used here are only locally defined, but for the sake of simplicity we drop this kind of technical detail for the time being. For a full  description of the basic details see \cite{CGM07, CGL08}.

A nonautonomous system of first-order ordinary differential equations in a manifold $N$ is represented
by a $t$-dependent vector field $X=X(t,x)$ on such a manifold. The  system of differential equations
associated with the $t$-dependent vector field $X(t,x)$ is written in local coordinates as
$$\frac{{\d}x^i}{{\d}t}=X^i(t,x)\,,\qquad i=1,\ldots,n=\dim\, N,
$$
where $X(t,x)=\sum_{i=1}^nX^i(t,x)\partial/\partial{x^i}$. The conditions for this system ensuring  that it admits
a superposition rule, i.e. there exists an open $U\subset N^{m+1}$ and a map
$\Phi:U\subset N^{(m+1)}\to N$ such that its general solution
can be written as $$x(t)=\Phi(x_{(1)}(t),
\ldots,x_{(m)}(t);k_1,\ldots,k_n),$$ where $\{x_{(a)}(t)\mid
a=1,\ldots,m\}$ is a family of particular solutions and
$k=(k_1,\ldots,k_n)$ is a set of   $n$ arbitrary  constants such that $$(x_{(1)},\ldots,x_{(m)},k_1,\ldots, k_n)\subset U,$$were
studied by S. Lie \cite{LS}. The necessary and sufficient condition is that the
associated  $t$-dependent vector field, $X(t)$, can be written as a
linear combination
\begin{equation}
X(t)=\sum_{\alpha =1}^r b_\alpha(t)\, X_{(\alpha )},
\label{Lievf}
\end{equation}
where the vector fields
 $\{X_{(\alpha)}\mid \alpha=1,\ldots,r\}$ are linearly independent vector
 fields, i.e. if $\lambda_1,\ldots, \lambda_r$  are 
real constants
 such that $\sum_{\alpha=1}^r\lambda_\alpha X_{(\alpha)}=0$, then $\lambda_1=\cdots=\lambda_r=0$,
 generating the so-called Vessiot-Guldberg Lie algebra $V$ of vector fields. The latter also means that there exist
 $r^3$ real numbers, $c_{\alpha\beta\gamma}$, such that
\begin{equation*}
[X_{(\alpha)},X_{(\beta)}]=\sum_{\gamma=1}^r c_{\alpha\beta\gamma}
X_{(\gamma)}\ ,\qquad \alpha,\beta=1,\ldots,r.
\end{equation*}

Any Lie system described by a $t$-dependent vector field on a manifold
 $N$, like (\ref{Lievf}), where the vector fields are complete
  and  close on a finite-dimensional Lie algebra determines a free, but maybe a discrete set of points, left action
$\Phi:G\times N\rightarrow N$ of a Lie group $G$ with Lie algebra $\mathfrak{g}\simeq V$ on the manifold $N$ describing the vector fields in $V$ as fundamental vector fields of such an action.

A solution of this system is represented by a curve $s\mapsto \gamma(s)$ in $N$ (integral curve) the tangent
vector of which $\dot \gamma$ at $t$ satisfies
\begin{equation}\label{e1}
\dot \gamma(t)= X(t, \gamma(t)).
\end{equation}
It is well-known that, at least for the smooth $X$ with which we work, for each $x_0$ there is a unique maximal
solution $\gamma_X^{x_0}(t)$  of system (\ref{e1}) with the initial value $x_0$, i.e. satisfying
$\gamma_X^{x_0}(0)=x_0$. The collection of all maximal solutions of the system (\ref{e1}) gives rise to a (local) generalised flow
$g^X$ on $N$. By a {\it generalised flow} $g$ on $N$ we understand a smooth $t$-dependent family $g_t$ of
 local diffeomorphisms on $N$, $g_t(x)=g(t,x)$, such that
$g_0=\text{id}_N$. The generalised flow $g^X$
induced by the $t$-dependent vector field $X$ is defined by
\begin{equation}\label{e3}
g^X(t,x_0)=\gamma_X^{x_0}(t)\,.
\end{equation}
Note that for $g=g^X$ equation (\ref{e3}) can be formally rewritten in the form
\begin{equation}\label{e4}
X(t)= X(t,x)=\dot g_t\circ g_t^{-1}\,.
\end{equation}

We observe that equation (\ref{e4}) in fact defines a one-to-one correspondence between generalised
flows and $t$-dependent vector fields. Any two generalised flows $g$ and $h$ can be composed: by definition $(g\circ h)_t=g_t\circ h_t$. As generalised flows correspond to $t$-dependent vector fields, this gives rise to an action  of a generalised flow $h$ on a $t$-dependent vector field $X$, giving rise to $h_\di X$ defined  by the equation
\begin{equation}
g^{h_\di X}=h\circ g^X\,. \label{e5}
\end{equation}
A more explicit form of this action is
\begin{equation}\label{e6}
(h_\di X)_t=\dot h_t\circ h_t^{-1}+(h_t)_*(X(t))\,,
\end{equation}
where $(h_t)_*$ is the standard action of diffeomorphisms on vector fields. These results can be summarised by means of the following theorem.

\begin{theorem}\label{t2} The equation (\ref{e6}) defines a natural action
of  generalised flows  on  $t$-dependent vector fields. This action is a group action in the sense that
$$(g\circ h)_\di X=g_\di(h_\di X).
$$
The integral curves of $h_\di X$ are of the form $h_t(\gamma(t))$ for $\gamma(t)$ being an arbitrary integral
curve for $X$.
\end{theorem}

We must notice that (\ref{e6}) and the theorem above still works even if $h$ is a one-parameter set of diffeomorphisms $h_t:N\rightarrow N$ with $h_0\neq Id$ and we can define the $t$-dependent vector field $h_{\di}X$ as the one with integral curves $h_t(\gamma(t))$, where $\gamma(t)$ is any integral curve for $X$.

Let us state the fundamental concepts of the theory of quasi-Lie schemes.

\begin{definition}{\rm Let $W$ and $V$ be nonnull finite-dimensional real vector spaces of vector fields on a manifold $N$.
We say that they form a {\it quasi-Lie scheme} $S(W,V)$ if they fulfil the conditions below:
\begin{enumerate}
\item $W$ is a vector subspace of $V$. \item $W$ is a Lie algebra of vector fields, i.e. $[W,W]\subset W$.
\item $W$ normalises $V$, i.e. $[W,V]\subset V$.
\end{enumerate}
If $V$ is a Lie algebra of vector fields $V$, we call the quasi-Lie scheme $S(V,V)$ simply a {\it Lie scheme}
$S(V)$.}
\end{definition}

There is the largest Lie subalgebra we can use as $W$ -- the normalizer of $V$ in $V$. Sometimes, however, it is useful to consider smaller Lie subalgebras $W$.

We say that a $t$-dependent vector field $X$ is in a quasi-Lie scheme $S(W,V)$ and write $X\in S(W,V)$ if
$X$ belongs to $V$ on its domain, i.e. $X(t)\in V$.

Now, given a quasi-Lie scheme $S(W,V)$ which we call sometimes simply a scheme, we may consider the group,
$\mathcal{G}(W)$, of generalised flows associated with $W$.
\begin{definition} {\rm We call the {\it group of the scheme} $S(W,V)$ the group $\mathcal{G}(W)$
of generalised flows corresponding to the $t$-dependent vector fields with values in $W$}.
\end{definition}
Given a scheme, $S(W,V)$, with $W$ a Lie algebra of complete
vector fields we can associate $W$ with a left action
$\Phi:G\times N\rightarrow N$, with $T_eG\simeq W$, describing
their elements as fundamental vector fields. Then the generalised
flows of $\mathcal{G}(W)$ are those of the form
$g_t(x)=\Phi(g(t),x)\equiv\Phi_{g(t)}(x)$ with $g(t)$ a curve in
$G$ with $g(0)=e$; see \cite{CGM07}.
\begin{proposition}\label{Main} {\rm Given a scheme $S(W,V)$, a $t$-dependent vector field $X\in S(W,V)$ and a generalised flow $g\in
\mathcal{G}(W)$, we get that $g_\di X\in S(W,V)$}.
\end{proposition}

We look for a set of $t$-dependent transformations containing $\mathcal{G}(W)$ satisfying the latter proposition.

\begin{lemma} Consider a scheme $S(W,V)$ with $W$ a Lie algebra of complete vector fields. Given an element $g\in\exp(\mathfrak{g})$ and a vector field $X\in S(W,V)$, then $\Phi_{g*}X\in S(W,V)$.
\end{lemma}
\begin{proof}

As $g\in\exp(\mathfrak{g})$, there exists an element ${\rm a}\in\mathfrak{g}$ such that $g=\exp({\rm a})$. Consider the curve $h:t\in[0,1]\rightarrow \exp(t\,{\rm a})\in G$. By means of the action $\Phi:G\times N\rightarrow N$ associated with the Lie algebra of vector fields $W$ of the scheme $S(W,V)$, the curve $h(t)$ induces the generalised flow $h^Y:(t,x)\in \mathbb{R}\times N\rightarrow h^Y_t(x)=\Phi(\exp(t\,{\rm a}),x)\in N$ for the vector field
$$
Y(x)=\dfrac{\d}{\d t}\bigg|_{t=0}h^Y_t(x)=\dfrac{\d}{\d t}\bigg|_{t=0}\Phi(\exp(t\,{\rm a}),x)
$$
and $Y\in W$. For each $t$ define the vector field $Z^{(0)}_t=h^Y_{t*}X$ to get
$$
(h_{t*}^YX)_x=X_x+\int^t_0\dfrac{\partial}{\partial s}Z_s^ {(0)}(x){\d s}=X_x+\int^t_0( h^Y_{s*}[Y,X])_x\d s.
$$
If we call $Z^{(1)}_t=h^Y_{t*}([Y,X])$ and apply the last formula to $[Y,X]$, we get
$$
\begin{aligned}
(h^Y_{s*}([Y,X]))_x&=[Y,X]_x+\int^{s}_0\dfrac{\partial}{\partial s'}Z^{(1)}_{s'}(x)\d s'\cr &=[Y,X]_x+\int^{s}_0(h^Y_{s'*}[Y,[Y,X]])_x\d s'.
\end{aligned}
$$
Defining $Z^{(k)}$ in an analogous way and applying all these results to the initial formula for $h^Y_{t*}X$ we obtain
\begin{multline*}
(h^Y_{t*}X)_x=X_x+[Y,X]_xt+\frac{1}{2}[Y,[Y,X]]_xt^2+\frac{1}{3!}[Y,[Y,[Y,X]]]_xt^3+\cdots.
\end{multline*}
By means of the properties of the scheme we obtain that each term belongs to the scheme, i.e.
$$[Y,[Y,\ldots,[Y,X]\ldots]]\in S(W,V),$$
and therefore
$$
\Phi_{g*}X=h^Y_{1*}X\in S(W,V).
$$
\end{proof}

\begin{proposition} Consider a scheme $S(W,V)$ with $W$ a Lie algebra of complete vector fields. Given a curve $g(t)\subset\exp(\mathfrak{g})$ and a $t$-dependent vector field $X(t)\in S(W,V)$, then $g_\di X\in S(W,V).$
\end{proposition}
\begin{proof} It has been shown that
$$
(g_\di X)_t=\dot g_t\circ g^{-1}_{t}+g_{t*}(X),
$$
but $\dot g_t\circ g^{-1}_{t}\in W\in S(W,V)$ and by means of the Lemma 1 we get that $g_{t*}X\in S(W,V)$ for each $t$. Hence we have $g_{\di}X\in S(W,V)$.
\end{proof}

\begin{definition}Given a scheme $S(W,V)$ we call the extended group of the scheme, ${\rm Ext}(W)$, the set of $t$-dependent transformations $\Phi_{g(t)}$ induced by curves $g(t)\subset\exp(\mathfrak{g})$ and the action $\Phi$ associated with the Lie algebra of complete vector fields $W$.

\end{definition}

From the last definition we can state the definition of {\rm quasi-Lie system} with respect to a scheme.
\begin{definition} {\rm Given a quasi-Lie scheme $S(W,V)$ and a $t$-dependent vector field $X\in S(W,V)$,
we say that $X$ is a {\it quasi-Lie system with respect to $S(W,V)$} if there exists a $t$-dependent transformation  $g\in
{\rm Ext}(W)$ and a Lie algebra of vector fields $V_0\subset V$ such that
$$
g_\di X\in S(V_0).
$$}
\end{definition}

\section{The Emden equation.}
\indent 

In this Section we approach from the perspective of
 the theory of quasi-Lie schemes the so-called Emden equations of the form
\begin{equation}\label{Emden}
\ddot x=a(t)\dot x+b(t)x^n,\quad n\ne 1.
\end{equation}
These equations can be  associated with the system of first-order differential equations
\begin{equation}\label{Emdsys}\left\{
\begin{array}{rcl}
\dot x&=&v,\\
\dot v&=&a(t)v+b(t)x^n.
\end{array}\right.
\end{equation}

This system was already studied in \cite{CGL08}
from the point of view of  the theory of quasi-Lie schemes.
 We summarise next the results of that paper and we use them
to obtain new properties: $t$-dependent constant of the motion by means of particular solutions, reducible particular cases of Emden equations etc.

Consider the real vector space, $V_{{\rm Emd}}$, spanned by the vector fields
\begin{equation*}
X_1=x\partial_{v},\quad X_2=x^n\partial_{v},\quad X_3=v\partial_{x},\quad
X_4=v\partial_{v},\quad X_5=x\partial_{x}.
\end{equation*}
The $t$-dependent vector field determining the dynamics of system
(\ref{Emdsys}) can be written as a linear combination
$$X(t)=a(t)X_4+X_3+b(t)X_2.
$$
Moreover the linear space $W_{{\rm Emd}}\subset V_{{\rm Emd}}$ spanned by the complete vector fields,
\begin{equation*}
Y_1=X_4=v\partial_{v},\quad Y_2=X_1=x\partial_{v},\quad Y_3=X_5=x\partial_{x},
\end{equation*}
  is a three-dimensional real Lie algebra of vector fields with respect to the
  ordinary Lie Bracket because these vector fields satisfy the relations
\begin{equation*}
\begin{aligned}
\left[Y_1,Y_2\right] &=-Y_2,\quad\left[Y_1,Y_3\right] &=0,\quad
\left[Y_2,Y_3\right] &=-Y_2.
\end{aligned}
\end{equation*}
Also $[W_{{\rm Emd}},V_{{\rm Emd}}]\subset V_{{\rm Emd}}$ because
\begin{equation*}
\begin{array}{lll}
\left[Y_1,X_2\right] =-X_2,& \left[Y_1,X_3\right] =X_3,&
\left[Y_2,X_2\right] =0,\cr \left[Y_2,X_3\right] =X_5-X_4,&
\left[Y_3,X_2\right] =nX_2,& \left[Y_3,X_3\right] =-X_3.
\end{array}
\end{equation*}
So we get a quasi-Lie scheme $S(W_{{\rm Emd}},V_{{\rm Emd}})$ which can be used to treat the Emden equations (\ref{Emdsys}). This suggests that we perform the $t$-dependent change of variables
associated with this
quasi-Lie scheme, namely,
\begin{equation}\label{transfor3}
\left\{
\begin{array}{rcl}
x&=&\gamma(t)x',\\
v&=&\beta(t)v'+\alpha(t)x',
\end{array}\right. \quad \gamma(t), \,\beta(t)> 0\,,\forall t,
\end{equation}
which transforms the original system into
{\small
\begin{equation}\label{transformed2}\left\{
\begin{aligned}\frac{{\d} x'}{{\d} t}&=\left(\frac{\alpha(t)}{\gamma(t)}-\frac{\dot \gamma(t)}{\gamma(t)}\right)x'+\frac{\beta(t)}{\gamma(t)}v',\\
\frac{{\d} v'}{{\d} t}&=\left(a(t)-\frac{\alpha(t)}{\gamma(t)}-\frac{\dot\beta(t)}{\beta(t)}\right)v'+\frac{\alpha(t)}{\beta(t)}\left(a(t)-\frac{\alpha(t)}{\gamma(t)}-\frac{\dot\alpha(t)}{\alpha(t)}+\frac{\dot\gamma(t)}{\gamma(t)}\right)x'\\
& \quad +\frac{b(t)\gamma^n(t)}{\beta(t)}x'{}^n.
\end{aligned}\right.
\end{equation}}

The key point of our method is choosing appropriate functions, $\alpha$,
$\beta $ and $\gamma$, in such a way that the system of differential equations (\ref{transformed2})  becomes
 a Lie system.

A possible way for the system (\ref{transformed2}) to be a Lie
system is to choose functions $\alpha, \beta$ and $\gamma$ such
that the latter system is determined by a $t$-dependent vector
field $X(t)=f(t)\bar X$, where $\bar X$ is a true vector field and
$f(t)$ is no vanishing function in the intervel of $t$ under study. As is
shown in next Section, this cannot always be  done and some
conditions must be imposed on the initial $t$-dependent functions,
$\alpha, \beta$ and $\gamma$, assuring such a transformation to
exist. These restrictions lead to integrability conditions.

Suppose for the time being that this is the case. Therefore the system (\ref{transformed2}) is
\begin{equation}\label{trans4}\left\{
\begin{aligned}
\frac{{\d} x'}{{\d} t}&=f(t)\left(c_{11}x'+c_{12}v'\right),\\
\frac{{\d} v'}{{\d} t}&=f(t)(c_{21}v'+c_xx'+c_{22}x'{}^n)
\end{aligned}\right.
\end{equation}
and it is determined by the $t$-dependent vector field
$$
X(t)=f(t)\bar X,
$$
with
$$
\bar{X}=(c_{11}x'+c_{12}v')\partial_{x'}+(c_{22}x'^n+c_xx'+c_{21}v')\partial_{v'}.
$$
Under the $t$-reparametrisation,
$$
\tau(t)=\int^t f(t')\d t',
$$
system (\ref{trans4}) is autonomous.
The new autonomous system of differential equations is determined by the vector
field $\bar X$ on ${\rm T}\mathbb{R}$ and therefore there exists a first
integral. This can be obtained by means of the method of characteristics,
which provides the characteristic curves where the first-integrals for
such a vector field
$\bar X$  are constant. These characteristic curves are determined by
$$
\frac{{\d} x'}{c_{11}x'+c_{12}v'}=\frac{{\d} v'}{c_{21}v'+c_xx'+c_{22}x'{}^n},
$$
which can be written as
\begin{equation}\label{eq21}
(c_{21}v'+c_xx'+c_{22}x'{}^n){\d} x'-(c_{11}x'+c_{12}v'){\d} v'=0.
\end{equation}
This expression can be straightforwardly integrated if
\begin{equation}\label{CInt}
\partial_{v'}(c_{21}v'+c_xx'+c_{22}x'{}^n)=-\partial_{x'}(c_{11}x'+c_{12}v')\Longrightarrow c_{21}=-c_{11}.
\end{equation}
Under this condition  we obtain the first integral for (\ref{eq21}), namely
\begin{equation}\label{IntOfMot}
I=-c_{12}\frac{v'{}^2}{2}+c_x\frac{x'{}^2}{2}+c_{21}v'x'+c_{22}\frac{x'{}^{n+1}}{n+1}.
\end{equation}
Finally, if we write the latter expression in terms of the initial variables $x,v$ and $t$, we get a constant  of the motion for the initial differential equation.

When  we do not want to impose condition (\ref{CInt}), we could also integrate equation (\ref{eq21}) by means of an integrating
factor,
 namely, we look for a function, $\mu(x',v')$, such that
\begin{equation*}
\partial_{v'}\left(\mu(c_{21}v'+c_xx'+c_{22}x'^n)\right)=\partial_{x'}(-\mu(c_{11}x'+c_{12}v')).
\end{equation*}
Thus the integrating factor satisfies the partial differential equation
$$
\pd{\mu}{v'}(c_{21}v'+c_xx'+c_{22}x'^n)+\pd{\mu}{x'}(c_{11}x'+c_{12}v')=-\mu(c_{11}+c_{21}).
$$
If $c_{11}+c_{21}=0$, the integral factor can be fixed to be $\mu=1$ and we get
the latter first integral (\ref{IntOfMot}). On the other hand, if
$c_{11}+c_{21}\neq 0$, we can still look for a solution for the partial
differential equation for $\mu$ and obtain a new first  integral.

\section{$t$-dependent constants of the motion and particular solutions.}
\indent

Our aim is to  show that the knowledge of a certain particular solution of the Emden
equation satisfying a certain condition allows us to transform  it into a Lie system and to obtain a
$t$-dependent constant of
 the motion. Even if some results related to this $t$-dependent constants of the motion can be found in the literature,  here we recover them from a new point of view.

If we restrict ourselves to the case $\alpha(t)=0$ in the system of differential equation
(\ref{transformed2}), it reduces to
\begin{equation}\label{eq8}\left\{
\begin{aligned}
\frac{{\d} x'}{{\d} t}&=-\frac{\dot \gamma(t)}{\gamma(t)}x'+\frac{\beta(t)}{\gamma(t)}v',\\
\frac{{\d} v'}{{\d} t}&=\left(a(t)-\frac{\dot\beta(t)}{\beta(t)}\right)v'+\frac{b(t)\gamma^n(t)}{\beta(t)}x'{}^n.
\end{aligned}\right.
\end{equation}

In order to transform the original Emden--Fowler differential equation
 into a Lie
system by means of our quasi-Lie scheme, we try writing the transformed
differential equation in the form
\begin{equation}\label{eq10}
\left\{
\begin{aligned}
\frac{{\d} x'}{{\d} t}&=f(t)\left(c_{11}x'+c_{12}v'\right),\\
\frac{{\d} v'}{{\d} t}&=f(t)\left(c_{22}x'{}^n+c_{21}v'\right),
\end{aligned}\right.
\end{equation}
where the $c_{ij}$ are constants. This system of differential equations can be
reduced to an autonomous one because under the $t$-dependent change of variables
$$
\tau(t)=\int^t f(t')\d t'
$$
the latter differential equation becomes
\begin{equation}\label{eq11}
\left\{
\begin{aligned}
\frac{{\d} x'}{{\d}\tau}&=c_{11}x'+c_{12}v',\\
\frac{{\d} v'}{{\d}\tau}&=c_{22}x'{}^n+c_{21}v'.
\end{aligned}\right.
\end{equation}
In order for system (\ref{eq8}) to be like system (\ref{eq10}) we look for functions $\alpha$, $\beta$ and
$\gamma$ satisfying the conditions,
\begin{equation}\label{Relations}
\left\{
\begin{aligned}
f(t)\,c_{11}&=-\frac{\dot\gamma(t)}{\gamma(t)}, \qquad &f(t)\,c_{12}&=\frac{\beta(t)}{\gamma(t)},\\
f(t)\,c_{22}&= b(t)\frac{\gamma^n(t)}{\beta(t)}, \qquad &
f(t)\,c_{21}&=a(t)-\frac{\dot\beta(t)}{\beta(t)}.
\end{aligned} \right.
\end{equation}
The conditions of the first line lead to
\begin{equation}\label{bet}
\beta(t)=-\frac{c_{12}}{c_{11}}\dot\gamma(t),
\end{equation}
and using this equation in the last relation we obtain
\begin{equation}\label{funo}
f(t)=\frac{a(t)}{c_{21}}-\frac{1}{c_{21}}\frac{\ddot\gamma(t)}{\dot\gamma(t)}.
\end{equation}
On the other hand, from the three first relations in (\ref{Relations}) we get
\begin{equation}\label{fdos}
f(t)=-\frac{b(t)c_{11}}{c_{22}c_{12}}\frac{\gamma^n(t)}{\dot\gamma(t)}.
\end{equation}

The equality of the right hand sides of (\ref{funo}) and (\ref{fdos}) leads to
the following equation
 for the function $\gamma$:
\begin{equation*}
\ddot\gamma=a(t)\dot\gamma+\frac{c_{11}c_{21}}{c_{22}c_{12}}b(t)\gamma^n.
\end{equation*}
Suppose that we make the choice, with $c_{21}=-c_{11}$ as indicated in (\ref{CInt}),
\begin{equation}\label{cschoice}
c_{22}=-1,\quad c_{11}=1,\quad c_{21}=-1,\quad c_{12}=1
\end{equation}
and thus $(c_{11}c_{22})/(c_{21}c_{12})=1$. Therefore we find that $\gamma$ must be a
solution of the initial equation (\ref{Emden}). In other words, if we suppose
 that a particular solution $x_p(t)$ of the Emden equation is known, we can
 choose $\gamma(t)=x_p(t)$.
Then, according to  the expression (\ref{bet}) and our previous choice
(\ref{cschoice}), the corresponding function $\beta$ turns out to be
$$\beta(t)=-\dot x_p(t).
$$
Finally, in view of conditions (\ref{Relations}), we get that
$$
\frac{-\dot\gamma(t)}{c_{11}\gamma(t)}=b(t)\frac{\gamma^ n(t)}{c_{22}\beta(t)}
$$
and taking into account our choice (\ref{cschoice}) and $\gamma(t)=x_p(t)$, we obtain the condition satisfied by the particular solution:
\begin{equation}\label{IntCond2}
x^{n+1}_p(t)={\dot x}^2_p(t).
\end{equation}
The system of differential equations (\ref{eq10}) for such a choice
(\ref{cschoice}) of the constants
$\{c_{ij}\,|\,i,j=1,2\}$ is the equation for the integrals curves for the $t$-dependent vector field
$$
X(t)=f(t)\left(\left(x'+\,v'\right)\partial_{x'}-\left(v'+\,{x'}^n\right)
\partial_{v'}\right).
$$
The method of the characteristics can be used to find the following
 first-integral for this vector field and, in view of (\ref{IntOfMot}), we get
\begin{equation*}
\left\{\begin{aligned}
I(x',v')&=\frac{1}{n+1}x'{}^{n+1}+\frac{1}{2}v'{}^2+x'v',\qquad &n&\notin \{ -1,1\},\\
I(x',v')&={\rm log}\,x'+\frac{1}{2}v'{}^2+x'v',\qquad &n&= -1,
\end{aligned}\right.
\end{equation*}
and, if we express this integral of motion in terms of the initial variables and
$t$, we obtain a, as far as we know, new $t$-dependent constant of the motion for the initial
Emden equation
\begin{equation}\label{Integral}
\left\{\begin{aligned}
I(t,x,v)&=\frac{x^{n+1}}{(n+1)x_p^{n+1}(t)}+\frac{v^2}{2\dot x_p^2(t)}-\frac{xv}{x_p(t)\dot x_p(t)},\quad &n&\notin\{ -1,1\},\\
I(t,x,v)&={\rm log}\left(\frac{x}{x_p(t)}\right)+\frac{v^2}{2\dot x_p^2(t)}-\frac{xv}{x_p(t)\dot x_p(t)},\quad &n&= -1.
\end{aligned}\right.
\end{equation}
So the knowledge of a particular solution for the Emden equation holding (\ref{IntCond2}) enables us
first to
obtain a constant of the motion and then to reduce the initial Emden equation into a Lie
system. Thus all Emden equations with such particular solutions are quasi-Lie
 systems with respect to the above mentioned  scheme, the applicability of which
 depends on a prior knowledge of such a particular solution.

\section{Applications of particular solutions to Emden equations.}\label{APS}
\indent 

This Section is devoted to  illustrate the theory with some particular instances of
Emden
 equations for which one is able to find a particular solution satisfying an integrability condition in an easy way
 and  use
is made of such a solution in order   to obtain the corresponding  $t$-dependent constant of the motion.
 In some cases the so obtained constant
  can be found in the literature, but they are found here from a new
  systematic procedure.

We start with a particular case of the Lane-Emden equation
\begin{equation}
\ddot x=-\frac{2}{t}\dot x-x^5.\label{Emdenpart}
\end{equation}
The most general
Lane-Emden equation is generally written as
$$
\ddot x=-\frac{2}{t}\dot x+f(x)
$$
and the example here  considered  corresponds to $f(x)=-x^n, \,
n\ne 1$, which is  one of the most interesting cases, together
with that of $f(x)=-e^{-\beta x}$. Equation (\ref{Emdenpart})
appears in the study of the thermal behavior of a spherical cloud
of gas \cite{KMM} and  also
 in  astrophysical applications. A particular solution for (\ref {Emdenpart}) satisfying (\ref{IntCond2}) is
 $x_p(t)=(2t)^{-1/2}$. If we substitute this expression for $x_p(t)$  and
the corresponding one for $\dot x_p(t)$ into the $t$-dependent constant of the motion
(\ref{Integral}), we get that
\begin{equation}\label{CMEF}
I'(t,x,v)=\frac{4t^3x^{6}}{3}+4t^ 3v^2+4t^ 2xv
\end{equation}
is a $t$-dependent constant of the motion proportional to (\ref{Integral}) and
also proportional to the $t$-dependent constants of the motion found in \cite{CGL08,Lo77,SB80}.

We study from this new perspective other Emden equations investigated in \cite{Le85}. Consider the particular instance
$$
\ddot x=-\frac{5}{t+K}\dot x-x^2.
$$
A particular solution for this Emden equation satisfying (\ref{IntCond2}) is
$$x_p(t)=\frac{4}{(t+K)^2}.
$$
In this case a $t$-dependent constant of the motion is
$$
I'(t,x,v)=\frac 13{x^3(t+K)^6}+\frac{1}{2}v^2(t+K)^6+2\,x\,v(t+K)^5,
$$
which is proportional to the one found by Leach in \cite{Le85}.

Now another Emden equation found in \cite{Le85},
$$
\ddot x=-\frac{3}{2(t+K)}\dot x-x^9,
$$
admits the particular solution
$$
x_p(t)=\frac{1}{\sqrt{2}(t+K)^{1/4}},$$
which satisfies (\ref{IntCond2}). The corresponding $t$-dependent constant of the motion is given by
$$
I'(t,x,v)=(K+t)^{3/2}(10(K+t)v^2+5 vx+2(K+t)x^{10})
$$
which is proportional to that given in \cite{Le85}.

As another example of Emden equation we can consider
$$
\ddot x=-\frac{5}{3(t+K)}\dot x-x^7,
$$
which admits as a particular solution
$$
x_p(t)=\frac{1}{3^{1/3}(t+K)^{1/3}},$$
which satisfies (\ref{IntCond2}) and leads to the $t$-dependent constant of the motion
$$
I'(t,x,v)=(K+t)^{5/3}(12(K+t)v^2+8 vx+3x^8(K+t)).
$$

Finally we apply our development to obtain a $t$-dependent constant of the motion for the Emden equation
\begin{equation}\label{eq}
\ddot x=-\frac{1}{K_1+K_3t}\dot x-x^n
\end{equation}
with
$$
K_3= \frac{n-1}{n+3}.$$
We can find a particular solution of the form
$$
x_p(t)=\frac{K_2}{(K_1+K_3t)^\nu}, \quad \nu\ne 0.
$$
In order $x_p(t)$ to be a particular solution we must hold the following relation
$$
\frac{(\nu+1)\nu K_2K_3^2}{(K_1+K_3t)^{\nu+2}}=\frac{\nu K_2K_3}{(K_1+K_3t)^{\nu+2}}-\frac{K_2^n}{(K_1+K_3t)^{n\nu}}
$$
and thus
$$
\nu+2=n\nu\qquad {\rm and}\qquad \nu(\nu+1)K_3^2K_2=\nu K_2 K_3-K_2^n.
$$
From these equations we get
$$
\nu=\frac{2}{n-1},\qquad K_2^{n-1}=\frac{2^2}{(n+3)^2}.
$$
Under these conditions it can be easily verified that $\dot x_p^2(t)=x_p^ {n+1}(t)$. Thus a $t$-dependent constant of the motion is
\begin{multline}\label{IntFinal}
I'(t,x,v)=(K_1+K_3t)^{2(n+1)/(n-1)}\left(\frac{x^{n+1}}{n+1}+\frac{v^2}{2}\right)+ \\+(K_1+K_3t)^{(n+3)/(n-1)}\frac{2vx}{n+3}, \qquad\qquad\end{multline}
which can be found also in \cite{Le85}.


Another advantage of our method is that it allows us to obtain Emden equations
admitting a previously fixed   $t$-dependent constant of the motion.

Suppose we want to construct an Emden equation with a previously
chosen particular solution, $x_p(t)$, satisfying ${\dot
x}^2_p(t)=x^{n+1}_p(t)$ for certain $n\in\mathbb{R}-\{1,-1\}$. We
can integrate this equation to get all possible particular
solutions which can be used by means of our method, i.e.
$$
x_p(t)=\left(K+\frac{1-n}{2}t\right)^{-\frac{2}{n-1}}.
$$
We consider functions $a(t)$ and $b(t)$ such that
$$
\ddot x_p=a(t)\dot x_p+b(t)x_p^n.
$$
For the sake of simplicity we can suppose that $b(t)=-1$. Then we get
$$
a(t)=\frac{\ddot x_p+x_p^n}{\dot x_p}.
$$
If we substitute in this expression for $a(t)$ the chosen particular solution,
we obtain
$$
a(t)=\frac{3+n}{2(K+\frac{1-n}{2}t)}.
$$
which leads to an Emden equation equivalent to (\ref{eq})  and the $t$-dependent constant of the motion for this equation is again (\ref{IntFinal}). In this way we recover the cases studied in this Section.

\section{The Kummer-Liouville transformation for a general Emden-Fowler equation.}\label{KL}
\indent 

The general form of the Emden--Fowler equation considered nowadays is
\begin{equation}\label{GEFeq}
\ddot x+p(t)\dot x+q(t)x=r(t)x^n.
\end{equation}
This generalisation arises naturally as a consequence of our scheme. This
latter second-order differential equation is associated with the system of first-order differential equations
\begin{equation}\label{FordGEFeq}
\left\{\begin{aligned}
\dot x&=v,\\
\dot v&=-p(t)v-q(t)x+r(t)x^n,
\end{aligned}\right.
\end{equation}
which is the system for the determination of  the integral curves for the $t$-dependent vector field
$$
X(t)=-p(t)X_4-q(t)X_1+r(t)X_2+X_3.
$$
This $t$-dependent vector field is a more general case than the one studied in previous Sections.
Under the set of transformations (\ref{transfor3})  the initial system
(\ref{FordGEFeq})
becomes the new system
{\small
\begin{equation*}\left\{
\begin{aligned}\frac{{\d} x'}{{\d} t}&=\left(\frac{\alpha(t)}{\gamma(t)}-\frac{\dot \gamma(t)}{\gamma(t)}\right)x'+\frac{\beta(t)}{\gamma(t)}v',\\
\frac{{\d} v'}{{\d} t}&=\left(-p(t)-\frac{\alpha(t)}{\gamma(t)}-\frac{\dot\beta(t)}{\beta(t)}\right)v'+
\frac{\alpha(t)}{\beta(t)}\left(-p(t)-\frac{\alpha(t)}{\gamma(t)}-\frac{\dot\alpha(t)}{\alpha(t)}+\right.\\ &\left.+\frac{\dot\gamma(t)}{\gamma(t)}-q(t)\frac{\gamma(t)}{\alpha(t)}\right)x'+\frac{r(t)\gamma^n(t)}{\beta(t)}x'{}^n.
\end{aligned}\right.
\end{equation*}}
If we choose $\alpha=\dot \gamma$,  the system reduces to
{\small
\begin{equation*}\left\{
\begin{aligned}\frac{{\d} x'}{{\d} t}&=\frac{\beta(t)}{\gamma(t)}v',\\
\frac{{\d} v'}{{\d} t}&=\left(-p(t)-\frac{\dot \gamma(t)}{\gamma(t)}-\frac{\dot\beta(t)}{\beta(t)}\right)v'+\frac{\dot \gamma(t)}{\beta(t)}\left(-p(t)-\frac{\ddot\gamma(t)}{\dot\gamma(t)}-q(t)\frac{\gamma(t)}{\dot\gamma(t)}\right)x'\\&+\frac{r(t)\gamma^n(t)}{\beta(t)}x'{}^n.
\end{aligned}\right.
\end{equation*}}
When the function $\gamma(t)$ is chosen to be such that $\ddot \gamma=-q(t)\gamma
-p(t)\dot\gamma$,
i.e. $\gamma$ is a solution of the associated linear equation, we obtain
{\small
\begin{equation}
\left\{
\begin{aligned}
\frac{{\d} x'}{{\d} t}&=\frac{\beta(t)}{\gamma(t)}v',\\
\frac{{\d} v'}{{\d} t}&=\left(-p(t)-\frac{\dot \gamma(t)}{\gamma(t)}-\frac{\dot \beta(t)}{\beta(t)}\right)v'+\frac{r(t)\gamma^n(t)}{\beta(t)}x'{}^n.
\end{aligned}\right.
\end{equation}}
Finally, if the function $\beta(t)$ is such that
$$
-p(t)-\frac{\dot \gamma(t)}{\gamma(t)}-\frac{\dot \beta(t)}{\beta(t)}=0,
$$
we obtain
{\small
\begin{equation}
\left\{
\begin{aligned}
\frac{{\d} x'}{{\d} t}&=\frac{\beta(t)}{\gamma(t)}v',\\
\frac{{\d} v'}{{\d} t}&=\frac{r(t)\gamma^{n}(t)}{\beta(t)}x'{}^n,
\end{aligned}\right.
\end{equation}}which is related to the second-order differential equation
$$
\frac{{\d}^2x'}{{\d}\tau ^2}=r(t)\frac{\gamma^{n+1}(t)}{\beta^2(t)}x'{}^n,$$
with
$$
\tau{(t)}=\int^t\frac{\beta(t')}{\gamma(t')}{\d} t'.
$$
 The new form of the differential equation is called the canonical form of the
 generalised Emden-Fowler equation.

This fact is obtained  in the previous literature by means of an appropriate
 Kummer-Liouville transformation, but here we obtain it as a straightforward application of the properties of transformation of quasi-Lie schemes thereby underscoring the theoretical explanation of such a Kummer-Liouville transformation.

\section{Constants of the motion for systems of Emden-Fowler equations.}
\indent

In this Section we show that under certain assumptions for the $t$-dependent
coefficients $a(t)$ and $b(t)$  the original Emden equation can be reduced
to a Lie system and then we can obtain a first integral  which
provides  us with a $t$-dependent constant of the motion for the original system.

In fact consider the system of first-order differential equations
(\ref{transformed2}). This system describes all the systems of
differential equations that can be obtained by means of the set of
$t$-dependent transformations we got through the scheme
$S(W_{Emd},V_{Emd})$. We recall that the $t$-dependent change of
variable which we use to relate the Emden equation (\ref{Emdsys})
with the latter system of differential equation is
(\ref{transfor3}). As in previous papers about this topic we try
to relate the initial system of differential equations to a Lie
system determined by a $t$-dependent vector field of the form
$X'(t)=f(t)\bar X$ and we suppose $f(t)$ to be non-vanishing
in the interval we study. So the system of differential equations
determining the integrals curves for this $t$-dependent vector
field is a Lie system and we can use the theory of Lie systems to
analyse its properties.

As a first example we can consider that we just use the set of transformations with $\gamma(t)=1$ and $\alpha(t)=0$. In this case system
(\ref{transfor3}) is
\begin{equation*}\left\{
\begin{aligned}\frac{{\d} x'}{{\d} t}&=\beta(t)v'\\
\frac{{\d} v'}{{\d} t}&=\left(a(t)-\frac{\dot\beta(t)}{\beta(t)}\right)v'+\frac{b(t)}{\beta(t)}x'{}^n.
\end{aligned}\right.
\end{equation*}
We fix $\beta(t)$ to  be such that
$$
a(t)-\frac{\dot\beta(t)}{\beta(t)}=0,
$$
i.e., $\beta(t)$ is (proportional to)
$$
\beta(t)=\exp\left(\int^ta(t')\d t'\right).
$$
Therefore we get
\begin{equation*}\left\{
\begin{aligned}\frac{{\d} x'}{{\d} t}&=\exp\left(\int^ta(t')\d t'\right)v',\\
\frac{{\d} v'}{{\d} t}&=b(t)\exp\left(-\int^ta(t')\d t'\right)x'{}^n.
\end{aligned}\right.
\end{equation*}
In order to get the last system of differential equations to describe the
integral curves for a $t$-dependent vector field, $X'(t,x)=f(t)\bar X(x)$, for
a given  function $a(t)$ a necessary and sufficient condition is
$$
b(t)\exp\left(-2\int^ta(t')\d t'\right)=K,
$$
with $K$ being a  real constant. Under this assumption the last system becomes
\begin{equation*}\left\{
\begin{aligned}\frac{{\d} x'}{{\d} t}&=\exp\left(\int^ta(t')\d t'\right)v',\\
\frac{{\d} v'}{{\d} t}&=\exp\left(\int^ta(t')\d t'\right)K x'{}^n.
\end{aligned}\right.
\end{equation*}
We introduce the $t$-reparametrisation
$$
\tau(t)=\int^t\exp\left(\int^{t'}a(t'')\d t''\right)\d t'
$$
and the latter system becomes
\begin{equation*}\left\{
\begin{aligned}\frac{{\d} x'}{{\d}\tau}&=v',\\
\frac{{\d} v'}{{\d}\tau}&=K x'{}^n,
\end{aligned}\right.
\end{equation*}
which admits a first integral
$$
I=\frac 12v'{}^2-K\frac{x'{}^{n+1}}{n+1}.
$$
In terms of the initial variables the corresponding $t$-dependent constant  of the motion is
$$
I=\exp\left(-2\int^ta(t')\d t'\right)\left(\frac 12{\dot x}^2-b(t)\frac{x^{n+1}}{n+1}\right),
$$
which is similar to that found in \cite{BV91}.

Suppose that we restrict the transformations (\ref{transfor3}) to the
case $\alpha(t)=0$. In this case the system of first-order differential equations (\ref{transformed2}) becomes
\begin{equation*}\left\{
\begin{aligned}\frac{{\d} x'}{{\d} t}&=-\frac{\dot \gamma(t)}{\gamma(t)}x'+\frac{\beta(t)}{\gamma(t)}v',\\
\frac{{\d} v'}{{\d} t}&=\left(a(t)-\frac{\dot\beta(t)}{\beta(t)}\right)v'+\frac{b(t)\gamma^n(t)}{\beta(t)}x'{}^n.
\end{aligned}\right.
\end{equation*}
In order for this system of differential equations to determine the integral curves for a $t$-dependent vector field of the form $X'(t)=f(t)\bar X$ we need that
\begin{equation}\label{Rel}
\left\{\begin{aligned}
c_{11}f(t)&=-\frac{\dot \gamma(t)}{\gamma(t)},\quad &c_{12}f(t)&=\frac{\beta(t)}{\gamma(t)},\\
c_{21}f(t)&=a(t)-\frac{\dot\beta(t)}{\beta(t)}, \quad &c_{22}f(t)&=\frac{b(t)\gamma^n(t)}{\beta(t)}.\\
\end{aligned}\right.
\end{equation}

From these relations, or more precisely from those of the first row, we get $f(t)$ as
$$
f(t)=-\frac{1}{c_{11}}\frac{\dot \gamma(t)}{\gamma(t)}=\frac{1}{c_{12}}\frac{\beta(t)}{\gamma(t)}
$$
and therefore
$$
\dot \gamma(t)=-\frac{c_{11}}{c_{12}}\beta(t).
$$
We choose  $c_{11}=-1$ and $c_{12}=1$ so that
\begin{equation}\label{beta}
\beta(t)=\dot \gamma(t).
\end{equation}
In view of this and using the third and second relations from (\ref{Rel}) we get
$$
\frac{c_{21}}{c_{12}}\frac{\beta(t)}{\gamma(t)}=a(t)-\frac{\dot\beta(t)}{\beta(t)}
$$
and thus, as a consequence of  (\ref{beta}), the last differential equation becomes
$$
\frac{c_{21}}{c_{12}}\frac{\dot\gamma(t)}{\gamma(t)}=a(t)-\frac{\ddot\gamma(t)}{\dot\gamma(t)}
$$
and, as $c_{12}=1$ and fixing $c_{21}=1$, we obtain
$$
\frac{\d}{\d t}\log(\dot\gamma\gamma)=a(t),
$$
which can be rewritten as
$$
\frac{1}{2}\frac{\d}{\d t}\gamma^2(t)=\exp\left(\int^ta(t')\d t'\right).
$$
Hence we have
$$
\gamma(t)=\sqrt{2\int^t\exp\left(\int^{t'}a(t'')\d t''\right)\d t'}
$$
and in view of (\ref{beta})
$$
\beta(t)=\frac{1}{\sqrt{2\int^t\exp\left(\int^{t'}a(t'')\d t''\right)\d t'}}\exp\left(\int^{t}a(t')\d t'\right).
$$

So far we have used only three of the four relations we found. The fourth and second
relations lead  to the integrability condition: there exists
a constant $c_{22}=K$ such that
$$
K\frac{\beta(t)}{\gamma(t)}=\frac{b(t)\gamma^n(t)}{\beta(t)}.
$$
Therefore, using the above  expressions for $\gamma(t)$ and $\beta(t)$, we get
\begin{equation}\label{Inte}
b(t)\exp\left(-2\int^ta(t)\d t'\right)\left(2\int^t\exp\left(\int^{t'}a(t'')\d t''\right)\right)^{(n+3)/2}=K.
\end{equation}

So under this assumption we have connected the initial Emden equation with the Lie system,
\begin{equation*}\left\{
\begin{aligned}\frac{{\d} x'}{{\d} t}&=f(t)(-x'+v'),\\
\frac{{\d} v'}{{\d} t}&=f(t)(v'+Kx'{}^n),
\end{aligned}\right.
\end{equation*}
and then  the method of characteristics shows that it  admits the first integral
$$
I'=-\frac{1}{2}v'^2+\frac{K}{n+1}x'{}^{n+1}+v'x'.
$$
In terms of the initial variables the corresponding constant of the motion is
\begin{multline}
I=\left(\frac 12\dot x^2-\frac{b(t)}{n+1}x^{n+1}\right)\exp\left(-2\int^ta(t')\d t'\right)\int^t\exp\left(\int^{t'}a(t'')\d t''\right)\d t'\\
-\frac{1}{2}x\dot x \exp\left(-\int^t a(t')\d t'\right)
\end{multline}
and in this way we recover the result found in \cite{BV91}. If we now consider
the particular case $n=-3$ we get that the integrability condition (\ref{Inte})
implies that there is a constant $K$ such that
\begin{equation*}
b(t)\exp\left(-2\int^ta(t)\d t'\right)=K,
\end{equation*}
and the corresponding $t$-dependent constant  of the motion is then given by
\begin{multline*}
I=\left(\frac 12\dot x^2+\frac{b(t)}{2}x^{-2}\right)\exp\left(-2\int^ta(t')\d t'\right)\int^t\exp\left(\int^{t'}a(t'')\d t''\right)\d t'\\
-\frac{1}{2}x\dot x \exp\left(-\int^t a(t')\d t'\right),\qquad\qquad
\end{multline*}
which is equivalent to that found in \cite{BV91}.
\section{A physical example.}\indent 

In order to motivate the physical applications of the theory of
quasi-Lie schemes to Emden-Fowler equations, we here develop a
detailed particular application of our methods to the Emden-Fowler
equations of the form
\begin{equation}\label{Pex}
\ddot x+\frac{2}{t}\dot x+x^n=0,\qquad n\neq 1.
\end{equation}
The case $n=1$ is not dealt with in this work because quasi-Lie schemes are not
needed to treat it. In this particular instance, the second-order differential
equation (\ref{Pex}) is linear and the introduction of the new  variable $v=\dot x$ transforms it into a nonautonomous linear system of first-order differential equations admitting a superposition rule. Thus (\ref{Pex}) is a SODE Lie system \cite{CLR08SIGMA} and the theory of quasi-Lie schemes reduces in this case to the usual theory of Lie systems.

The equations (\ref{Pex}) initially appeared in the study of the
stellar structure to analyse the configuration of spherical clouds of gas. These equations are
also used in gas dynamics, fluid dynamics and more recently in
relativistic mechanics, nuclear physics and chemical reacting
systems \cite{Wong76a}.

We must point out that the analysis of the solutions for equations
(\ref{Pex}) is particularly important. Special interest have those
non-oscillatory solutions with positive zeros. These zeros are
related to equilibrium states of a fluid in a spherical
distribution of density and under mutual attraction of particles.
Some results about the existence of these solutions can be found
in \cite{Wong76a, YO07,KSY88}. Furthermore, those solutions $x(t)$
with initial conditions $x(0)=1$ and $\dot x(0)=0$ are specially
important in astrophysics \cite{CC07,La70}. In what follows, we
analyse some transformation properties of the equations
(\ref{Pex}) and  derive some new particular solutions. These
solutions are interesting because they can provide information
about stability of fluids in a spherical configuration or they can be used to check out
methods to analyse particular solutions
 \cite{YO07,KSY88}.

It is known that (\ref{Pex}) admits certain analytic solutions for
the cases $n=0,1$ and $n=5$, e.g  the case $n=0$ has a particular solution $x(t)=-t^2/6$, the case $n=1$,  the particular one $x(t)=\sin t/t$ and the case $n=5$ has the particular solution $x_p(t)=(2t)^{-1/2}$, see \cite{Wong76a} and previous Sections. From the latter solution, we can recover another very well-known solution of the equation (\ref{Pex}). Suppose we want to get the
particular solution $x(t)$ for the equation (\ref{Pex}) with
$x(0)=1$ and $\dot x(0)=0$. Thus, using this values at $t=0$ in
the constant of the motion (\ref{CMEF}) obtained by means of the particular solution $x_p(t)=(2t)^{-1/2}$, we get that such a
constant is zero along the solution $x(t)$ with the previous initial
conditions and we have
\begin{equation*}
\frac{4t^3x^{6}}{3}+4t^ 3\dot x^2+4t^ 2x\dot x=0,
\end{equation*}
from which we obtain 
\begin{equation*}
 \frac{dx}{dt}=\frac{x}{2t}\left(-1+\sqrt{1-
\frac 43 x^4t^2}\right).
\end{equation*}
We make the change of variables $x^2=m$ and $\tau=t^2$ in order to
simplify the above differential equation and to get
\begin{equation*}
\frac{dm}{d\tau}=\frac{m}{2\tau}\left(-1+\sqrt{1-
\frac 43 m^2\tau}\right).
\end{equation*}
A new change of variables $n=1/m$ transforms the above equation into
$$
\frac{dn}{d\tau}=\frac{1}{2\tau}\left(n-\sqrt{n^2-
\frac {4\tau}{3}}\right).
$$
In order to remove the squared term in the previous differential
equation, at least for a certain particular solution, we perform
the change of variables $n=u+\tau/3$ and we obtain
$$
\frac{du}{d\tau}+\frac{1}{3}=\frac{1}{2\tau}\left(u+\frac{\tau}3-\sqrt{u^2+2\frac{\tau}3u+\frac{\tau^2}{3^2}-
\frac {4\tau}{3}}\right).
$$
It is easy to check that the latter has a particular solution
$u=1$. Then inverting all previous changes of variables we get
the well-known particular solution for (\ref{Pex}) with $x(0)=1$
and $\dot x(0)=1$ is given by
$$
x(t)=\left(1+\frac{t^2}{3}\right)^{-1/2}.
$$
We show that the knowledge of this particular solution allows
us to obtain a way to get a family of exact solutions for the
equation (\ref{Pex}) with $n=5$. In this way we illustrate how
the knowledge of a particular solution can be used to obtain
information about the general solution. From a physical point of
view, this fact enables us to
investigate by means of particular solutions the existence  in the example 
of bounded oscillatory solutions the zeros of which determine
 equilibrium states of a fluid with spherical distribution of density and under mutual attraction of its particles \cite{CGS61,Sa40}.

Recall that the particular solution $x_p(t)=(2t)^{-1/2}$ allowed
us to transform the first-order system of differential equations
related to (\ref{Pex}) to the one describing the integral curves
for the $t$-dependent vector field
$$
X(t)=f(t)\left((x'+v')\partial_{x'}-(v'+x'^5)\partial_{v'}\right).
$$
Lie's theorem shows that any system determined by a $t$-dependent
vector field on a manifold $N$ of the form $X(t,p)=f(t)\widehat
X(p)$, with $p\in \N$, is a Lie system related to a
onedimensional Lie algebra of vector fields. Moreover, the theory
of Lie systems explains how to get a superposition rule for this
Lie system \cite{CGM07}. Indeed, we have to consider $m+1$ copies
of the system. In this case, the solutions of the total system are
integral curves for a $t$-dependent vector field of the form $\widetilde X(t)=f(t)\widehat X$ with
$$
\widetilde X(t)=f(t)\left(\sum_{a=0}^m (x_{(a)}'+v_{(a)}')\partial_{x'_{(a)}}-(v'_{(a)}+x_{(a)}'^5)\partial_{ v_{(a)}'}\right).
$$
The method developed in \cite{CGM07} explains that we have to consider $m$ in
such a way that $\pi_*\widehat X$  be linearly independent with $\pi$ the canonical projection $\pi: ({\rm T}\mathbb{R})^{n(m+1)}\rightarrow({\rm T}\mathbb{R})^{nm}$ onto the last $nm$ variables.  Then the first-integrals for $\widehat X$ provide as the superposition rule. In our case, for $m=1$ the vector field $\pi_*\widehat X$ is linearly independent and the first-integrals for $\widehat X$ lead to the superposition rule. More explicitly, the 
vector field
$$
\widehat X=\sum_{a=0}^1 (x_{(a)}'+v_{(a)}')\partial_{x'_{(a)}}-(v'_{(a)}+x_{(a)}'^5)\partial_{v_{(a)}'},
$$
admits locally three first-integrals allowing us to obtain the superposition
rule. The method of characteristics provides us with two of such first-integrals
\begin{equation*}
\begin{aligned}
I_i(x_{(i)}',v_{(i)}')&=\frac{1}{6}x^{'6}_{(i)}+\frac{1}{2}v^{'2}_{(i)}+x_{(i)}'v_{(i)}',\qquad
i=0,1.
\end{aligned}
\end{equation*}
In order to obtain an analytic expression for a third (partial)
first-integral, we restrict ourselves to the submanifold $$\mathcal{C}=\{(x',v')\in ({\rm T}\mathbb{R})^2\,|\,I_0(x',v')=I_1(x',v')=0\},$$ 
with $(x',v')=(x'_{(0)},v'_{(0)},x_{(1)}',v_{(1)}')$. As is shown below, this case
is enough general to obtain new exact solutions for (\ref{Pex}).
Under the assumed restrictions, a third first-integral $K$ on $\mathcal{C}$, that is $\widehat XK|_{\mathcal{C}}=0$, is
\begin{equation}
K=\frac{{x'}^2_{(0)}}{{x'}^2_{(1)}}\left(\frac{1+\sqrt{1-\frac{{x'}_{(1)}^4}3}}{1+\sqrt{1-\frac{{x'}_{(0)}^4}3}}\right).
\end{equation}
From the latter expression we can work out the value of $x'_{(0)}$
in terms of $x'_{(1)}$ and the constant $K$
$$
x'_{(0)}=\sqrt{\frac{6K{x'}_{(1)}^2\left(1-\sqrt{1-\frac{{x'}_{(1)}^4}3}+K^2\left(1+\sqrt{1-\frac{{x'}_{(1)}^4}3}\right)\right)}{12K^2+(1-K^2)^2{x'_{(1)}}^4}},
$$
and taking into account that $x'_{(i)}=\sqrt{2t}\,x_i$, with $i=0,1$,
we get what we call a partial $t$-dependent superposition
rule for the equations (\ref{Pex}):
$$
x_0=\sqrt{\frac{6Kx_1^2\left(1-\sqrt{1-\frac{4t^2x_1^4}3}+K^2\left(1+\sqrt{1-\frac{4t^2x_1^4}3}\right)\right)}{12K^2+(1-K^2)^24t^2x_1^4}}.
$$
It can be seen that  this $t$-dependent partial superposition rule
does not generate all solutions for the Emden-Fowler equation
(\ref{Pex}). The reason is that in order to get an algebraic
expression for such a partial superposition we had to restrict
ourselves to some kind of solutions. The name of partial is
related to the partial superposition rule concept \cite{CGM07}.
These partial superposition rules do not enable us to obtain the
general solution, but a family of solutions.

If we take into account that $x_1(t)=(1+t^2/3)^{-1/2}$ is a
particular solution for the equation (\ref{Pex}) with $n=5$, we
can use such a solution in our $t$-dependent partial superposition
rule to get the $K$ parametrised family of solutions
$$
x_0(t)=\sqrt{\frac
32}\sqrt{\frac{K\left(3+t^2-|-3+t^2|+K^2(3+t^2+|-3+t^2|)\right)}{(3K^2+t^2)(3+K^2t^2)}}.
$$
In particular, for $K=1$ we recover $x_0(t)=(1+t^2/3)^{-1/2}$
and for $K=0$ we get $x_0(t)=0$. We can see that for any
$K$ we get that $\dot x_0(0)=0$.

This $t$-dependent partial superposition rule can be used to
analyse the solutions of the equation (\ref{Pex}). In particular,
it can be seen that in the limit $t\rightarrow\infty$ all
solutions $x_0(t)$ tend to zero, i.e. $\lim_{t\rightarrow\infty}
x_0(t)=0$.

 Finally, we proved in Section \ref{KL} that the
system (\ref{FordGEFeq}) related to an Emden-Fowler equation of
the form (\ref{GEFeq}) can be transformed by means of a
$t$-dependent change of variables (\ref{transfor3}) with
$\alpha(t)=\dot \gamma(t)$ and $\beta$ and $\gamma$ two
$t$-dependent functions holding the differential equations
$$
\ddot \gamma=-q(t)\gamma -p(t)\dot\gamma\qquad {\rm and}\qquad
-p(t)-\frac{\dot \gamma(t)}{\gamma(t)}-\frac{\dot
\beta(t)}{\beta(t)}=0,
$$
into the system
\begin{equation*}
\left\{
\begin{aligned}
\frac{{\d} x'}{{\d} t}&=\frac{\beta(t)}{\gamma(t)}v',\\
\frac{{\d} v'}{{\d} t}&=\frac{r(t)\gamma^{n}(t)}{\beta(t)}x'{}^n.
\end{aligned}\right.
\end{equation*}
For the case of Emden-Fowler equations (\ref{Pex}) we have that $p(t)=2/t$, $q(t)=0$ and $r(t)=1$. Thus, we get that
$$
\ddot \gamma= -\frac{2}{t}\dot\gamma \qquad {\rm and}\qquad
-\frac{2}{t}-\frac{\dot \gamma(t)}{\gamma(t)}-\frac{\dot
\beta(t)}{\beta(t)}=0.
$$
Hence, $\gamma(t)=\beta(t)=t^{-1}$ and we obtain
\begin{equation*}
\left\{
\begin{aligned}
\frac{{\d} x'}{{\d} t}&=v',\\
\frac{{\d} v'}{{\d} t}&=t^{1-n}x'{}^n.
\end{aligned}\right.
\end{equation*}
Therefore, we can see that for any solution 
$(x'(t),v'(t))$ of the above system, the function $x'(t)$ satisfies the second-order
differential equation
$$
\frac{{\d}^2x'}{{\d}t^2}=t^{1-n}x'^n,
$$
and the change of variable $x=t^{-1}x'$ transforms the equation (\ref{Pex})
into the above differential equation. Even whether this result can
be found in \cite{Wong76a}, we here provide a new approach to it.
\section{Conclusions and Outlook.}\indent

We have applied the theory of quasi-Lie schemes to treat Emden
equations. We have shown that the knowledge of a particular
solution satisfying a certain condition enables us to transform a
given Emden equation into a Lie system. As a consequence  we get a family of 
Emden equations which are quasi-Lie systems with respect to the
quasi-Lie scheme we have used. This fact provides a method to
obtain $t$-dependent constants of the motion. We have offered a
method to construct Emden equations with a previously fixed
particular solution related to a $t$-dependent constant of the
motion. We have also explained the Kummer-Liouville transformation
from the point of view of the theory of Lie schemes. Next, we have
applied a method used in previous papers about quasi-Lie schemes to
obtain $t$-dependent constants of the motion for certain families
of Emden equations. Finally, in order to motivate the physical
applications of our methods, we have analysed certain cases of
Emden-Fowler equations. For such cases, some properties of
transformation have been derived. Additionally, for an 
interesting particular Emden equation, we have recovered a particular
solution and through such a solution, and by means of a
$t$-dependent partial superposition rule for this Emden equation,
a family of, as far as we know, new solutions has been derived.

As it was stated in \cite{CGL08}, the field of applications of quasi-Lie schemes
is very broad. Here we have presented a new application of this concept and we hope that more results
can be recovered or obtained by means of our methods, e.g. new
$t$-dependent partial superposition rules to obtain particular
exact solutions from particular ones.

\section*{Acknowledgements.}

 Partial financial support by research projects MTM2006-10531 and E24/1 (DGA)
 are acknowledged. JdL also acknowledges a F.P.U. grant from  Ministerio de
 Educaci\'on y Ciencia.  PGLL thanks the University of KwaZulu-Natal and
 the National Research Foundation of South Africa for their continued support.
 The opinions expressed in this paper should not be construed to be those of the organisation.


\end{document}